# Iterative Nonlocal Residual Elasticity

## Mohamed Shaat*

*Mechanical Engineering Department, Zagazig University, Zagazig 44511, Egypt*

**Abstract**

Motivated by the existing complications of finding solutions of Eringen's nonlocal model, an alternative model is developed here. The new formulation of the nonlocal elasticity is centered upon expressing the dynamic equilibrium requirements based on a nonlocal residual stress field. This new nonlocal elasticity is explained from the lattice mechanics and continuum mechanics points of view. Boundary value problems obtained based on the new nonlocal elasticity are solved following a proposed iterative procedure. This iterative procedure is centered upon correcting the solution of the classical field problem for the nonlocal residual field of the elastic domain. Convergence analyses are presented to show the convergence of the iterative procedure to the solution of the nonlocal field problem. The iterative procedure is an integrated part of the proposed nonlocal elasticity. Therefore, the newly developed nonlocal elasticity is give the name "iterative nonlocal residual elasticity".

## 1. Introduction

In classical mechanics, a body consists of an infinite number of particles each of which is a mass point. Each particle undergoes interactions only with the nearest particles (neighbor particles). Whereas the interactions between non-neighbor particles are weaker than interactions between neighbor particles, these non-neighbor interactions are exist and may contribute to the continuum's mechanics in certain occasions. Therefore, the nonlocal theory was developed to propose a generalized theory, which models the particle exhibits interactions with its neighbors and non-neighbors.

According to Eringen [1,2], the dynamic equilibrium of a solid elastic body is conditional by the global balance of its body forces, external surface tractions, and inertia forces. The nonlocal theory, as early proposed by Eringen, postulates the condition of the global balance, thus [2]:

-------------------------------------------------------------------------------------------------

*Corresponding author.
E-mail addresses: shaat@nmsu.edu; shaatscience@yahoo.com (M. Shaat).

$$t_{ji,j} + F_i + \mathcal{F}_i = \rho \frac{\partial^2 u_i}{\partial t^2} \tag{1}$$

where $\rho$ is the mass density, $u_i$ is the displacement field of a material particle, and $t$ is the time. $F_i$ is a local-type body force, which is conjugate to neighbor interactions. $\mathcal{F}_i$ is a nonlocal body force. The nonlocal body force ($\mathcal{F}_i$) is a residual field, which generates due to non-neighbor interactions. This nonlocal residual body force yields to the condition [1,2]:

$$\int\limits_V \mathcal{F}_i \, dV = 0 \tag{2}$$

where $V$ is the volume of the elastic body. For a homogeneous-isotropic-linear elastic material, $\mathcal{F}_i = 0$ [1,2].

The balance laws of nonlocal continua (equation (1)) depend on a total stress field ($t_{ij}$), which sums neighbor interactions and non-neighbor interactions at a specific point $\boldsymbol{x}$ belongs to the elastic body. This total stress field ($t_{ij}$) was expressed as a functional of the deformation gradients of all points of the elastic body [1,2]. Thus, for a homogeneous-isotropic-linear elastic material, the constitutive equations were determined based on some invariance and thermodynamics requirements with the form [1,2]:

$$t_{ij}(\boldsymbol{x}) = \int\limits_V \lambda'(|\boldsymbol{x}' - \boldsymbol{x}|)\varepsilon_{rr}(\boldsymbol{x}')\delta_{ij} + 2\mu'(|\boldsymbol{x}' - \boldsymbol{x}|)\varepsilon_{ij}(\boldsymbol{x}') \, dV(\boldsymbol{x}') \tag{3}$$

where $\varepsilon_{ij}$ is the infinitesimal strain. $\lambda'$ and $\mu'$ are Lame moduli, which are functionals of $|\boldsymbol{x}' - \boldsymbol{x}|$. The integration over the elastic domain, $V$, was involved to collect all non-neighbor interactions.

With Eringen's manipulation of the nonlocal theory, the balance laws are identical to those of the classical continuum mechanics [1,2]. However, the constitutive law is different where the total stress field ($t_{ij}$) was introduced to model neighbor and non-neighbor interactions via integral operators (equation (3)). This total stress field is commonly known as "nonlocal stress field".

Complications of finding solutions of nonlocal elasticity problems have been early discussed [3]. According to equation (3), the field equation of the nonlocal problem is an integropartial differential equation. Analytical solutions for integropartial differential equations are difficult to be determined, especially for mixed boundary value problems [3]. This motivated Eringen [3] to formulate the nonlocal elasticity in terms of singular differential equations. By assuming that (i) the Lame moduli exhibit the same attenuation with the distance $|\boldsymbol{x}' - \boldsymbol{x}|$ according to (ii) a Green's function of a differential operator with constant coefficients, the nonlocal field problem was formulated as a partial differential equation [3]:

$$\sigma_{ji,j} + \mathcal{L}\left(F_i - \rho \frac{\partial^2 u_i}{\partial t^2}\right) = 0 \tag{4}$$

where $\mathcal{L}$ is the differential operator as defined by Eringen [3], *i.e.* $\mathcal{L}t_{ij} = \sigma_{ij}$. $\sigma_{ij}$ is the local stress of the classical elasticity.

Whereas solutions can be easily obtained using the differential form of the nonlocal elasticity (equation (4)), the ability of this form to secret exact solutions, which are consistent with the original form that depend on the integral operator is controversial. Discrepancies were observed between solutions of the differential nonlocal field problems and integral nonlocal field problems [4-7]. Details on the paradoxes of the differential nonlocal elasticity can be found in [4-13].

Moreover, the "ill-posedness" of Eringen's nonlocal elasticity was discussed in [14]. It was revealed that the transformation of the integral constitutive law (equation (3)) into a differential nonlocal constitutive law secretes inconsistencies between the natural boundary conditions of the nonlocal equilibrium and the boundary conditions of the constitutive model generated upon the transformation [14].

An example of the paradox and inconsistency of the differential nonlocal elasticity can be exhibited by considering a nonlocal problem of an elastic body under a body force equals its inertia force, in particular $F_i - \rho \frac{\partial^2 u_i}{\partial t^2} = 0$. An example of the latter case is a nonlocal static problem with a zero body force. For this case, equation (4) reduces to:

$$\sigma_{ji,j} = 0 \qquad\qquad\qquad (5)$$

which gives the stress field identical to that of the classical elasticity with no dependency on the nonlocal character of the body. This is a clear violation of the concept of the nonlocal mechanics.

The illustrated example and the paradoxes and inconsistencies expressed in previous studies imply that the transformation of the integral nonlocal constitutive law (3) into a differential nonlocal constitutive law as proposed by Eringen [3] may be not convenient to express nonlocal boundary value problems. However, with the deep inspection of the situation, one can realize the fact that the transformation into the differential constitutive law proposed in [3] is mathematically consistent. So, the question is: what are the reasons behind the paradoxes of Eringen's nonlocal elasticity?

An attempt to provide an answer to this question was recently expressed in [14]. Whereas this study defined new insights on the ill-posedness of nonlocal field problems formed based on the "pure nonlocal elasticity", it was claimed that a two-phase local/nonlocal mixture constitutive law overcomes the paradoxes and inconsistencies of Eringen's nonlocal elasticity. However, a comment was represented in [15] on this false statement to demonstrate that the ill-posedness is an inherited trait of Eringen's constitutive laws and there is no way to overcome it.

Motivated by the aforementioned complications and debates regarding Eringen's nonlocal elasticity presented in equations (1)-(4), an alternative form of the nonlocal theory is developed in this study. This new form of the nonlocal elasticity depends on splitting the total nonlocal stress field $(t_{ij})$ into a local stress ($\sigma_{ij}$) and a nonlocal residual stress ($\tau_{ij}$), *i.e.* $t_{ij} = \sigma_{ij} + \tau_{ij}$. The local stress ($\sigma_{ij}$) is a stress field that models the material stiffness due to neighbor (local) interactions. The nonlocal residual stress ($\tau_{ij}$) models the material stiffness due to non-neighbor (nonlocal) interactions. In the context of the new nonlocal elasticity, the balance laws are similar to those of the classical elasticity of an elastic continuum with a residual stress field. The proposed nonlocal theory is centered upon modifying the classical elasticity for an inherited nonlocal residual field. At the boundary of the elastic body, the surface tractions depend on both the local stress and the nonlocal residual stress.

The new manipulation of the nonlocal theory proposed here permits implementing an iterative procedure to extract solutions of nonlocal field problems. This iterative procedure is centered upon correcting the solution of the classical field problem for the nonlocal residual field of the elastic domain. In the context of the proposed approach, the classical field problem of an elastic domain with a pre-formed nonlocal residual stress is solved. Convergence analyses are presented to show the convergence of the iterative procedure to the solution of the nonlocal field problem formed based on the proposed nonlocal theory. It should be mentioned that the iterative procedure is an integrated part of the theory. Therefore, the newly developed form of the nonlocal theory will be give the name "iterative nonlocal residual elasticity".

## 2. Nonlocal Residual Elasticity

In the classical mechanics, a balance law is valid not only for the entire elastic body but also for each point belongs to the elastic domain [2]. However, in the nonlocal mechanics, the balance is achieved globally for the entire body. Thus, a nonlocal balance law is violated at a local point while satisfied globally because of the nonlocal interactions [2]. The nonlocal mechanics distinguishes between a local-type field and a nonlocal-type field. A nonlocal field (*e.g.* nonlocal stress field) is the sum of a local field, which accounts for the direct-neighbor interactions, and a nonlocal residual field. The nonlocal residual field is the sum of all the nonlocal interactions at a point, as shown in Fig. 1.

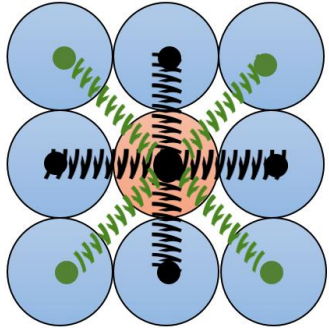

**Figure 1: Nonlocal lattice mechanics.** Dark interactions are direct-neighbor interactions while the green ones are non-neighbor interactions (only up to the next-nearest interaction).

## 2.1. Nonlocal mechanics of particles

To give more insights on the nonlocal elasticity developed here, the system of particles presented in Fig. 1 is considered. As shown in the figure, a set of interaction forces is generated between a particle ($n$) and its neighbor particles. In the classical mechanics, these direct interactions are exist. Each one of these interactions forms a local force between two neighbors. From lattice mechanics, the local force field, $\boldsymbol{F}$, between two neighbor particles ($n$ and $m$) can be defined as follows:

$$\boldsymbol{F}_{nm} = \boldsymbol{K} : (\boldsymbol{u}_n - \boldsymbol{u}_m) \tag{6}$$

where $\boldsymbol{K}$ is an equivalent stiffness of an interaction between two neighbors. $\boldsymbol{u}$ is the displacement field expressing the translational motion of the particle.

In addition to the direct neighbor interactions, interactions are generated between a particle ($n$) and all other non-neighbor particles. A non-neighbor interaction, $\boldsymbol{\mathcal{F}}$, depends on the distance between the two particles. Thus, the non-neighbor interaction between two non-neighbors ($n$ and $j$) is a functional of the distance between them $|\boldsymbol{x}_n - \boldsymbol{x}_j|$, as follows:

$$\boldsymbol{\mathcal{F}}_{nj} = \boldsymbol{k}_r(|\boldsymbol{x}_n - \boldsymbol{x}_j|) : (\boldsymbol{u}_n - \boldsymbol{u}_j) \tag{7}$$

where $\boldsymbol{k}_r(|\boldsymbol{x}_n - \boldsymbol{x}_j|)$ is a stiffness that is equivalent to the interaction between two non-neighbor particles.

Under the influence of the neighbor and non-neighbor interactions, the condition of the dynamic equilibrium of a particle ($n$) secretes the following balance equations:

$$\sum_{m=1}^{N_m} \boldsymbol{F}_{nm} + \sum_{j=1}^{N_j} \boldsymbol{\mathcal{F}}_{nj} = \frac{\partial^2}{\partial t^2}(\bar{m}\boldsymbol{u})$$

$$\sum_{m=1}^{N_m} (\mathbf{X} \times \boldsymbol{F}_{nm}) + \sum_{j=1}^{N_j} (\mathbf{X} \times \boldsymbol{\mathcal{F}}_{nj}) = \frac{\partial^2}{\partial t^2}(\mathbf{X} \times \bar{m}\boldsymbol{u}) \tag{8}$$

where $\mathbf{X} \times \boldsymbol{\xi}$ denotes a moment field. $N_m$ is the number of neighbors of particle ($n$) while $N_j$ is the number of its non-neighbor particles. $\bar{m}$ is the particle's mass.

We introduce in equations (6)-(8) a new form of the nonlocal theory in which the nonlocal model is formulated depending on the nonlocal residual field ($\sum_{j=1}^{N_j} \boldsymbol{\mathcal{F}}_{nj}$), which is the sum of all the non-neighbor (nonlocal) interactions of the particle under consideration. As a crucial requirement of nonlocal mechanics, the balance equations (8) are obtained from the global statement of equilibrium.

In Eringen's form of the nonlocal theory, the balance equations were formed depending on a total field that merges the nonlocal residual field with the local field. However, in the context of the proposed nonlocal elasticity, the nonlocal residual field is formed separately, and it is not merged with the local field. In addition, two types of constitutive equations are needed. In addition to the local-type constitutive equations (which are used to form the local field), constitutive equations are used to form the nonlocal residual field.

## 2.2. Nonlocal continuum mechanics: Balance laws

The global balance laws of a nonlocal continuum with a volume $V$ and bounded by a surface $S$ can be derived according to equation (8), as follows:

$$\int_V \boldsymbol{F} \, dV + \int_V \boldsymbol{\mathcal{F}} \, dV + \int_S \boldsymbol{T} \, dS = \int_V \rho \ddot{\boldsymbol{u}} \, dV \tag{9}$$

$$\int_V (\mathbf{X} \times \boldsymbol{F}) \, dV + \int_V (\mathbf{X} \times \boldsymbol{\mathcal{F}}) \, dV + \int_S (\mathbf{X} \times \boldsymbol{T}) \, dS = \int_V (\mathbf{X} \times \rho \ddot{\boldsymbol{u}}) \, dV \tag{10}$$

where $\rho$ is the mass density. $\boldsymbol{F}$ denotes the local body force. $\boldsymbol{\mathcal{F}}$ is a nonlocal residual body force. $\boldsymbol{T} \, dS$ denotes a surface traction.

Like the classical mechanics, the external surface traction ($\boldsymbol{T}$) is balanced with the body forces and the inertia of the elastic domain (equations (9) and (10)). However, in the proposed nonlocal elasticity, the surface traction $\boldsymbol{T}$ is formed depending on two tractions:

$$\boldsymbol{T} = \boldsymbol{n} \cdot \boldsymbol{\sigma} + \boldsymbol{n} \cdot \boldsymbol{\tau} \tag{11}$$

where $\boldsymbol{n}$ is the unit normal vector. $\boldsymbol{\sigma}$ is the local stress, and $\boldsymbol{\tau}$ is a nonlocal residual stress. The local stress ($\boldsymbol{\sigma}$) is conjugate to the direct neighbor interactions while the nonlocal residual stress ($\boldsymbol{\tau}$) is conjugate to the non-neighbor interactions. It should be observed that the stress tensors, $\boldsymbol{\sigma}$ and $\boldsymbol{\tau}$, are general tensors which have symmetric and skew-symmetric parts.

According to equation (11) and the divergence theorem, equations (9) and (10) can be rewritten as follows:

$$\boldsymbol{F} + \boldsymbol{\mathcal{F}} + \nabla \cdot \boldsymbol{\sigma} + \nabla \cdot \boldsymbol{\tau} = \rho \ddot{\boldsymbol{u}} \tag{12}$$

$$(\boldsymbol{\sigma} + \boldsymbol{\tau}) \times \boldsymbol{I} = 0 \tag{13}$$

Equation (13) indicates that the skew-symmetric parts of the local stress tensor, $\boldsymbol{\sigma}$, and the nonlocal residual stress tensor, $\boldsymbol{\tau}$, vanish. Therefore, the balance equations can be written in the form:

$$\boldsymbol{F} + \boldsymbol{\mathcal{F}} + \nabla \cdot \boldsymbol{\sigma}^{sym} + \nabla \cdot \boldsymbol{\tau}^{sym} = \rho \ddot{\boldsymbol{u}} \tag{14}$$

where the balance equations depend on the symmetric parts of the stress tensors, $\boldsymbol{\sigma}^{sym}$ and $\boldsymbol{\tau}^{sym}$. In the rest of the paper, the superscript '$sym$' is omitted for convenient writing.

## 2.3. Nonlocal continuum mechanics: Constitutive model

The balance laws of Eringen's model of the nonlocal linear elasticity depend on a total stress field that merges the local stress with the nonlocal residual stress. This total-nonlocal stress field was formed as a functional depending on the motions histories of all points of the body. However, the balance laws of the proposed nonlocal elasticity (equation (14)) depend on a nonlocal residual stress field, which is not merged with the local stress field. In this section, constitutive equations are derived for the nonlocal residual stress field.

Eringen expressed a postulation [1] that states, "*For the linear theory of nonlocal elastic materials, whose natural state is free of nonlocal effects, the nonlocal body force vanishes.*", *i.e.* $\mathcal{F} = 0$. In light of this postulation, the balance laws of a homogeneous-isotropic-linear elastic solid material can be written in the form:

$$\sigma_{ji,j} + \tau_{ji,j} + F_i = \rho \ddot{u}_i \tag{15}$$

where $\sigma_{ji} = \sigma_{ij}$ and $\tau_{ji} = \tau_{ij}$, which are the components of the local stress tensor and the nonlocal residual stress tensor, respectively.

For a linear elastic material, the infinitesimal strain is the fundamental measure of deformation:

$$\varepsilon_{ij} = \frac{1}{2}\left(u_{i,j} + u_{j,i}\right) \tag{16}$$

The local stress, $\sigma_{ji}$, has the same form as the stress field of the classical mechanics:

$$\sigma_{ji} = \lambda_0 \varepsilon_{rr}\delta_{ij} - 2\mu_0\varepsilon_{ij} \tag{17}$$

where $\lambda_0$ and $\mu_0$ are the Lame constants as defined in the classical mechanics.

Following Eringen's definition of nonlocal fields, the following constitutive equation can be proposed for the nonlocal stress residual ($\tau_{ij}$):

$$\tau_{ij}(\boldsymbol{x}) = \int_V \lambda_r(|\boldsymbol{x}' - \boldsymbol{x}|)\varepsilon_{rr}(\boldsymbol{x}')\delta_{ij} + 2\mu_r(|\boldsymbol{x}' - \boldsymbol{x}|)\varepsilon_{ij}(\boldsymbol{x}')\, dV(\boldsymbol{x}') \tag{18}$$

where $\lambda_r$ and $\mu_r$ are Lame moduli, which model the stiffness of a non-neighbor interaction. The magnitudes of these Lame moduli depend of the distance between two non-neighbor particles within the elastic domain ($|\boldsymbol{x}' - \boldsymbol{x}|$).

## 3. Boundary Value Problem and Solution Procedure: Iterative Nonlocal Residual Elasticity

The boundary value problem of the homogeneous-isotropic-linear elastic material formed based on the proposed nonlocal elasticity constitutes the following system of equations:

$$
\begin{aligned}
&1:\ \sigma_{ji,j}(\boldsymbol{x}) + \tau_{ji,j}(\boldsymbol{x}) + F_i(\boldsymbol{x}) = \rho \ddot{u}_i(\boldsymbol{x}) && \forall \boldsymbol{x} \in V \ \text{(Equation of Motion)} \\
&2:\ \sigma_{ji}(\boldsymbol{x})n_j + \tau_{ji}(\boldsymbol{x})n_j = T_i(\boldsymbol{x}) \ \text{ or } \ u_i(\boldsymbol{x}) = U_i(\boldsymbol{x}) && \forall \boldsymbol{x} \in S \ \text{(Boundary Conditions)} \\
&3:\ \sigma_{ji}(\boldsymbol{x}) = \lambda_0 \varepsilon_{rr}(\boldsymbol{x})\delta_{ij} + 2\mu_0 \varepsilon_{ij}(\boldsymbol{x}) && \forall \boldsymbol{x} \in V \ \text{(Constitutive Equations)} \\
&4:\ \tau_{ij}(\boldsymbol{x}) = \mathcal{R}\Big(\varepsilon_{ij}(\boldsymbol{x})\Big) && \forall \boldsymbol{x} \in V \ \text{(Constitutive Equations)} \\
&5:\ \varepsilon_{ij}(\boldsymbol{x}) = \tfrac{1}{2}\Big(u_{i,j}(\boldsymbol{x}) + u_{j,i}(\boldsymbol{x})\Big) && \forall \boldsymbol{x} \in V \ \text{(Kinematical Variables)}
\end{aligned}
$$

where

$$
\mathcal{R}\Big(\varepsilon_{ij}(\boldsymbol{x})\Big) = \int\limits_{V} \lambda_r(|\boldsymbol{x}' - \boldsymbol{x}|)\varepsilon_{rr}(\boldsymbol{x}')\delta_{ij} + 2\mu_r(|\boldsymbol{x}' - \boldsymbol{x}|)\varepsilon_{ij}(\boldsymbol{x}') \, dV(\boldsymbol{x}')
$$

where $U_i(\boldsymbol{x})$ is a prescribed displacement field at the boundary.

(19)

It should be mentioned that the boundary value problem (equation (19)) is similar to the one represented based on Eringen's nonlocal elasticity [1,2,16]. However, the new form of the nonlocal boundary value problem presented in equation (19) depends on splitting the total nonlocal stress field into a local stress field and a nonlocal residual stress field.

In this section, an iterative procedure is proposed to solve the nonlocal boundary value problem (equation (19)). With the implementation of the iterative procedure proposed here, the nonlocal field equation is converted into a local-type field equation with an imposed nonlocal residual field. In each iteration, the nonlocal residual stress is known. This nonlocal residual stress is formed based on the stress field obtained in a previous iteration. Thus, the implementation of this iterative procedure gives the merit of forming the nonlocal residual stress iteratively based on a pre-determined local-type field. In the context of this iterative procedure, a local-type boundary value problem is corrected for the nonlocal field and then solved.

The iterative procedure implemented in this study is pioneered from an iterative-finite element-based model of nonlocal elasticity proposed in [16] and implemented in [17,18] for beams and plates. In this study, a new iterative procedure is implemented to extract solutions of the proposed boundary value problem (equation (19)).

In an iteration $(k)$, the nonlocal boundary value problem (19) is converted into a local boundary value problem of an elastic body exposed to a residual stress field, as follows:

$$
\begin{aligned}
&1:\ \sigma_{ji,j}^{(k)}(\boldsymbol{x}) + \tau_{ji,j}^{(k)}(\boldsymbol{x}) + F_i(\boldsymbol{x}) = \rho \ddot{u}_i^{(k)}(\boldsymbol{x}) &&\forall \boldsymbol{x} \in V \ \ (\text{Equation of Motion})\\[4pt]
&2:\ \sigma_{ji}^{(k)}(\boldsymbol{x})n_j = T_i(\boldsymbol{x}) - \tau_{ji}^{(k)}(\boldsymbol{x})n_j \ \text{or}\ u_i^{(k)}(\boldsymbol{x}) = U_i(\boldsymbol{x}) \ \ \forall \boldsymbol{x} \in S \ \ (\text{Boundary Conditions})\\[4pt]
&3:\ \sigma_{ji}^{(k)}(\boldsymbol{x}) = \lambda_0 \varepsilon_{rr}^{(k)}(\boldsymbol{x})\delta_{ij} + 2\mu_0 \varepsilon_{ij}^{(k)}(\boldsymbol{x}) &&\forall \boldsymbol{x} \in V \ \ (\text{Constitutive Equations})\\[4pt]
&4:\ \tau_{ij}^{(k)}(\boldsymbol{x}) = \mathcal{R}\!\left(\varepsilon_{ij}^{(k-1)}(\boldsymbol{x})\right) &&\forall \boldsymbol{x} \in V \ \ (\text{Constitutive Equations})\\[4pt]
&5:\ \varepsilon_{ij}^{(k)}(\boldsymbol{x}) = \tfrac{1}{2}\!\left(u_{i,j}^{(k)}(\boldsymbol{x}) + u_{j,i}^{(k)}(\boldsymbol{x})\right) &&\forall \boldsymbol{x} \in V \ \ (\text{Kinematical Variables})
\end{aligned}
$$

where

$$
\mathcal{R}\!\left(\varepsilon_{ij}^{(k-1)}(\boldsymbol{x})\right) = \int_V \lambda_r(|\boldsymbol{x}' - \boldsymbol{x}|)\varepsilon_{rr}^{(k-1)}(\boldsymbol{x}')\delta_{ij} + 2\mu_r(|\boldsymbol{x}' - \boldsymbol{x}|)\varepsilon_{ij}^{(k-1)}(\boldsymbol{x}')\,dV(\boldsymbol{x}')
$$

(20)

Equations (20) form a local boundary value problem of an elastic domain exposed to a pre-defined residual stress field, $\tau_{ij}$. A procedure to derive a solution of the nonlocal boundary value problem (equation (20)) is given as follows:

1: **PROCEDURE:** in iteration $k$,
2:   (I) Nonlocal residual stress, $\tau_{ji}^{(k)}(\boldsymbol{x})$, formation:
3:     → *Recall $\varepsilon_{ij}^{(k-1)}(\boldsymbol{x})$, the determined strain field of the previous iteration $k-1$.*
4:     → *Form $\tau_{ji}^{(k)}(\boldsymbol{x})$ using equation (20)$_4$.*
5:     → *Save $\tau_{ji}^{(k)}(\boldsymbol{x})$.*
6:   (II) Solving the local boundary value problem:
7:     → *Recall $\tau_{ji}^{(k)}(\boldsymbol{x})$.*
8:     → *Form the local boundary value problem (equations (20)$_1$ and (20)$_2$).*
9:     → *Solve the local boundary value problem for $\sigma_{ji}^{(k)}(\boldsymbol{x})$ and $\varepsilon_{ij}^{(k)}(\boldsymbol{x})$.*
10:     → *Save $\sigma_{ji}^{(k)}(\boldsymbol{x})$ and $\varepsilon_{ij}^{(k)}(\boldsymbol{x})$.*
11: **End Procedure**

The proposed PROCEDURE presents a simple but effective approach to determine the solution of a nonlocal field problem formed based on the proposed iterative nonlocal residual elasticity. In an iteration $k$, the determined fields of a previous iteration $k-1$ ($\sigma_{ji}^{(k-1)}(\boldsymbol{x})$ and $\varepsilon_{ij}^{(k-1)}(\boldsymbol{x})$) are used to form the nonlocal residual stress, $\tau_{ji}^{(k)}(\boldsymbol{x})$, according to equation (20)$_4$. Then, the determined nonlocal residual stress is substituted into the local boundary value problem (22)$_1$ and (22)$_2$. This local boundary value problem is then solved for the fields $\sigma_{ji}^{(k)}(\boldsymbol{x})$ and $\varepsilon_{ij}^{(k)}(\boldsymbol{x})$. The procedure is then repeated where the solution converges to the nonlocal solution with the iterations.

It follows from the presented PROCEDURE that a solution is guaranteed in each iteration where the local boundary value problem is the one that is solved (a local boundary value problem is well-posed and admits a unique solution). The convergence of the obtained solution to the targeted nonlocal solution is constraint by the correct formation of the nonlocal residual stress. Indeed, with the new form of the nonlocal elasticity proposed here, the paradoxes and the inconsistencies associated with Eringen's nonlocal elasticity are effectively resolved. Examples of different boundary value problems are considered in the next section to demonstrate this fact.

## 4.  Application to Euler-Bernoulli Beams

In this section, a set of nonlocal elasticity problems of Euler-Bernoulli beams are solved using the proposed iterative nonlocal residual elasticity.

Because the nonlocal residual stress only models non-neighbor interactions, its Lame moduli ($\lambda_r$ and $\mu_r$) can be considered as follows:

$$\lambda_r(|\boldsymbol{x}|) = \lambda_0\big(\beta_\lambda(|\boldsymbol{x}|) - \delta(|\boldsymbol{x}|)\big)$$

$$\mu_r(|\boldsymbol{x}|) = \mu_0\left(\beta_\mu(|\boldsymbol{x}|) - \delta(|\boldsymbol{x}|)\right)$$

with

(21)

$$\beta_\lambda(|\boldsymbol{x}|) = \tfrac{1}{2\ell_\lambda}\exp\left(-\tfrac{|\boldsymbol{x}|}{\ell_\lambda}\right) \quad \text{and} \quad \beta_\mu(|\boldsymbol{x}|) = \tfrac{1}{2\ell_\mu}\exp\left(-\tfrac{|\boldsymbol{x}|}{\ell_\mu}\right)$$

where $\delta(|\boldsymbol{x}|)$ is a Dirac-delta function.

The substitution of equation (21) into equation (18) yields:

$$\tau_{ij}(\boldsymbol{x}) = \int\limits_V \left(\lambda_0\beta_\lambda(|\boldsymbol{x}' - \boldsymbol{x}|)\varepsilon_{rr}(\boldsymbol{x}')\delta_{ij} + 2\mu_0\beta_\mu(|\boldsymbol{x}' - \boldsymbol{x}|)\varepsilon_{ij}(\boldsymbol{x}')\right)dV(\boldsymbol{x}')$$
$$- \lambda_0\varepsilon_{rr}(\boldsymbol{x})\delta_{ij} - 2\mu_0\varepsilon_{ij}(\boldsymbol{x})$$

(22)

which is an explicit form of the nonlocal residual stress ($\tau_{ij}$). This constitutive law depends on $\beta_\lambda(|\boldsymbol{x}' - \boldsymbol{x}|)$ and $\beta_\mu(|\boldsymbol{x}' - \boldsymbol{x}|)$, which are two independent kernels that define nonlocal effects. These kernels depend on two independent nonlocal parameters $\ell_\lambda$ and $\ell_\mu$. As previously demonstrated by the author, a general form of the nonlocal theory with expanded applicability is the one that considers different nonlocal kernels for the different material coefficients [19].

According to equation (20) and equation (22), the boundary value problem of the elastostatic equilibrium of the bending of a beam subjected to a transverse load distribution $q(x)$ is derived as follows:

$$1: \frac{d^2 M^{(k)}(x)}{dx^2} + \frac{d^2 \mathcal{M}^{(k)}(x)}{dx^2} = q(x) \qquad\qquad \forall x \in V \quad \text{(Equilibrium Equation)}$$

$$2: \frac{dM^{(k)}(0 \text{ or } L)}{dx} = \bar{V} - \frac{d\mathcal{M}^{(k)}(0 \text{ or } L)}{dx} \;\; \text{or } w(0 \text{ or } L) = \bar{w}$$

$$3: M^{(k)}(0 \text{ or } L) = \bar{M} - \mathcal{M}^{(k)}(0 \text{ or } L) \;\; \text{or } \frac{dw(0 \text{ or } L)}{dx} = \overline{w_x}$$

$$\qquad\qquad\qquad\qquad\qquad\qquad\qquad\qquad \forall x \in S \quad \text{(Boundary Conditions)}$$

$$4: M^{(k)}(x) = -(\lambda_0 + 2\mu_0)\frac{bh^3}{12}\frac{d^2 w^{(k)}(x)}{dx^2}$$

$$5: \mathcal{M}^{(k)}(x) = \int_0^L \left( \xi_\lambda \beta_\lambda(|x'-x|) + \xi_\mu \beta_\mu(|x'-x|) \right) M^{(k-1)}(x')\, dx' - M^{(k-1)}(x)$$

$$\qquad\qquad\qquad\qquad\qquad\qquad\qquad \forall x \in V \quad \text{(Constitutive Equations)}$$

$$6: \varepsilon_{xx}^{(k)}(x) = -z \frac{d^2 w^{(k)}(x)}{dx^2} \qquad\qquad \forall x \in V \quad \text{(Kinematical Variables)}$$

where

$$\xi_\lambda = \frac{\lambda_0}{\lambda_0 + 2\mu_0} \quad \text{and} \quad \xi_\mu = \frac{2\mu_0}{\lambda_0 + 2\mu_0}$$

$$\beta_\lambda(|x'-x|) = \frac{1}{2\ell_\lambda}\exp\left(-\frac{|x'-x|}{\ell_\lambda}\right) \quad \text{and} \quad \beta_\mu(|x'-x|) = \frac{1}{2\ell_\mu}\exp\left(-\frac{|x'-x|}{\ell_\mu}\right)$$

(23)

where $M(x)$ is the local-type bending moment, and $\mathcal{M}(x)$ is the nonlocal residual bending moment. $w(x)$ is the beam deflection. $L$, $b$ and $h$ are the beam length, width and height, respectively. $\bar{V}$, $\bar{M}$, $\bar{w}$, and $\overline{w_x}$ are, respectively, prescribed shear force, bending moment, deflection, and slope at the boundary.

The proposed PROCEDURE is used to obtain the solution of the nonlocal boundary value problem (equation (23)). The procedure starts such that the nonlocal residual stress $\mathcal{M}(x)$ is zero. In this case, a solution of the boundary value problem (equations $(23)_1$, $(23)_2$, and $(23)_3$) gives the local solution of the bending field $M(x)$. Then, the obtained bending field ($M(x)$) is substituted into the constitutive law $(23)_5$ to form the nonlocal residual stress ($\mathcal{M}(x)$). Afterwards, $\mathcal{M}(x)$ is substituted into the boundary value problem (equations $(23)_1$, $(23)_2$, and $(23)_3$), which is then solved for the bending field $M(x)$. This process, when repeated, secretes the bending field $M(x)$ updated in each iteration. After a few iterations, both, the bending field $M(x)$ and the nonlocal residual bending field $\mathcal{M}(x)$ converge to their targeted solutions.

It should be mentioned that the bending field $M(x)$ is a local-type field, which accounts for the direct-neighbor interactions. This local-type field is adapted for the inclusion of the nonlocal residual fields of the elastic domain. The bending field $M(x)$ when added to the nonlocal residual bending $\mathcal{M}(x)$ forms a total bending field, which forms the constitutive law of Eringen's nonlocal elasticity, *i.e.*

$$M_{total}(x) = M(x) + \mathcal{M}(x) \qquad\qquad\qquad (24)$$

It should also be mentioned that the total bending field $M_{total}(x)$ formed by the sum of the bending fields $M(x)$ and $\mathcal{M}(x)$ is exactly the same as the one obtained from the equilibrium equation of Eringen's nonlocal elasticity. This total bending satisfies the requirement of the global balance.

Figures 2-4 show the nonlocal residual bending moment $\mathcal{M}(x)$, the curvature, and the deflection of cantilever, simply supported, and clamped-clamped beams. The boundary conditions for the different beam configurations are defined in each iteration $(k)$, as follows:

**Cantilever beam with a point load ($F$) at its free-end:**

$$\frac{dM^{(k)}(L)}{dx} = -F - \frac{d\mathcal{M}^{(k)}(L)}{dx}$$

$$M^{(k)}(L) = -\mathcal{M}^{(k)}(L)$$

$$w(0) = 0$$

$$\frac{dw(0)}{dx} = 0$$

**Simply supported beam under a uniform load ($q$):**

$$M^{(k)}(0) = -\mathcal{M}^{(k)}(0)$$

$$M^{(k)}(L) = -\mathcal{M}^{(k)}(L)$$

$$w(0) = 0$$

$$w(L) = 0$$

$$(25)$$

**Clamped-clamped beam under a uniform load ($q$):**

$$M^{(k)}(0) = \frac{qL^2}{12} - \mathcal{M}^{(k)}(0)$$

$$M^{(k)}(L) = \frac{qL^2}{12} - \mathcal{M}^{(k)}(L)$$

$$w(0) = 0$$

$$w(L) = 0$$

As one of the main drawbacks of Eringen's nonlocal elasticity, it secretes stress fields identical to those of the classical elasticity. This is because it only forms the total stress field, which is always identical to the stress field of the classical elasticity due to the equilibrium requirements. Thus, Eringen's nonlocal elasticity totally hides the role of the nonlocal residual field and gives no information about it. This leads to the paradoxes and inconsistencies previously discussed.

The proposed iterative nonlocal residual elasticity outweighs Eringen's nonlocal elasticity where it distinguishes between nonlocal residual fields, local-type fields, and total nonlocal fields. The proposed nonlocal elasticity can effectively form nonlocal residual fields (this is for the first time). In addition, it secretes a stress field dependent on the nonlocal character of the elastic domain. This stress field is the local-type stress field ($M(x)$).

Figures 2-4 show the effectiveness of the proposed iterative nonlocal residual elasticity to model nonlocal problems. In the first iteration, a local solution is obtained (the local solution is represented by broken-blue curves in Figures 2-4). After a few iterations, the solution converges to the targeted solution. The convergence is evident in Figures 2-4.

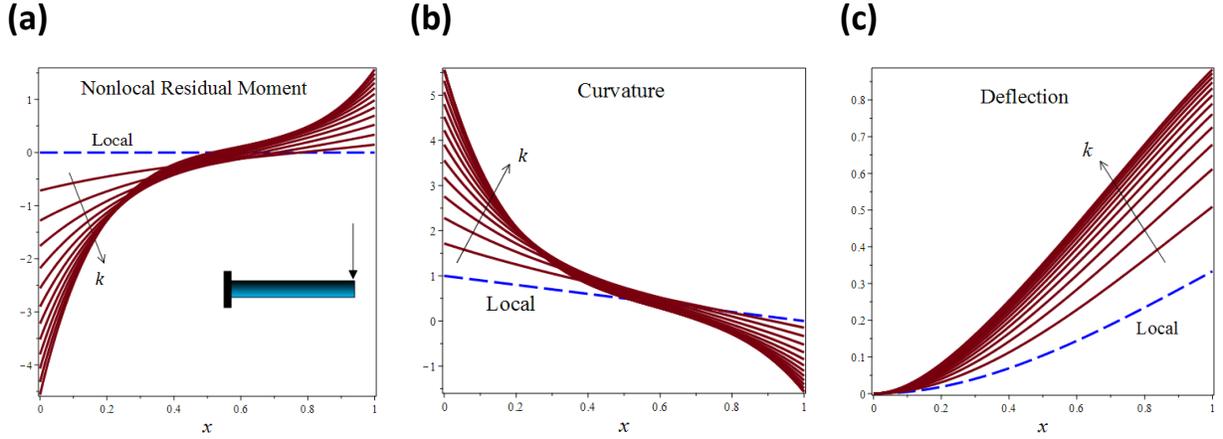

**Figure 2: Cantilever beam under a point load ($F = 1$):** (a) The nonlocal residual moment ($\mathcal{M}(x)$), (b) the curvature ($\frac{d^2w(x)}{dx^2} = \frac{M(x)}{(\lambda_0 + 2\mu_0)\frac{bh^3}{12}}$), and (c) the deflection ($w(x)$) at each iteration ($k$). The broken curves represent the local solutions. $\xi_\lambda = 1/3$, $\xi_\mu = 2/3$, $\ell_\lambda = \ell_\mu = 0.5$, $L = 1$.

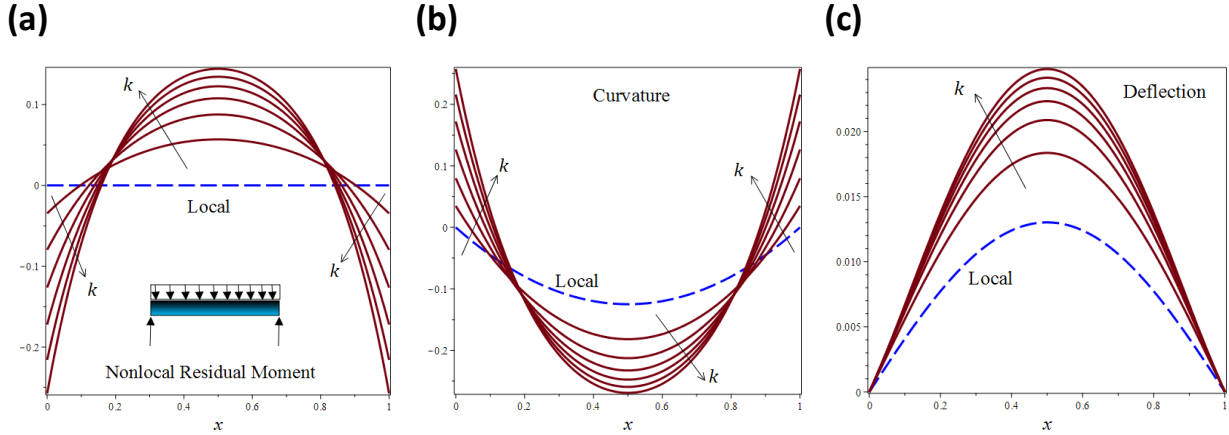

**Figure 3: Simply supported beam under a uniform load ($q = 1$):** (a) The nonlocal residual moment ($\mathcal{M}(x)$), (b) the curvature ($\frac{d^2w(x)}{dx^2} = \frac{M(x)}{(\lambda_0 + 2\mu_0)\frac{bh^3}{12}}$), and (c) the deflection ($w(x)$) at each iteration ($k$). The broken curves represent the local solutions. $\xi_\lambda = 1/3$, $\xi_\mu = 2/3$, $\ell_\lambda = \ell_\mu = 0.4$, $L = 1$.

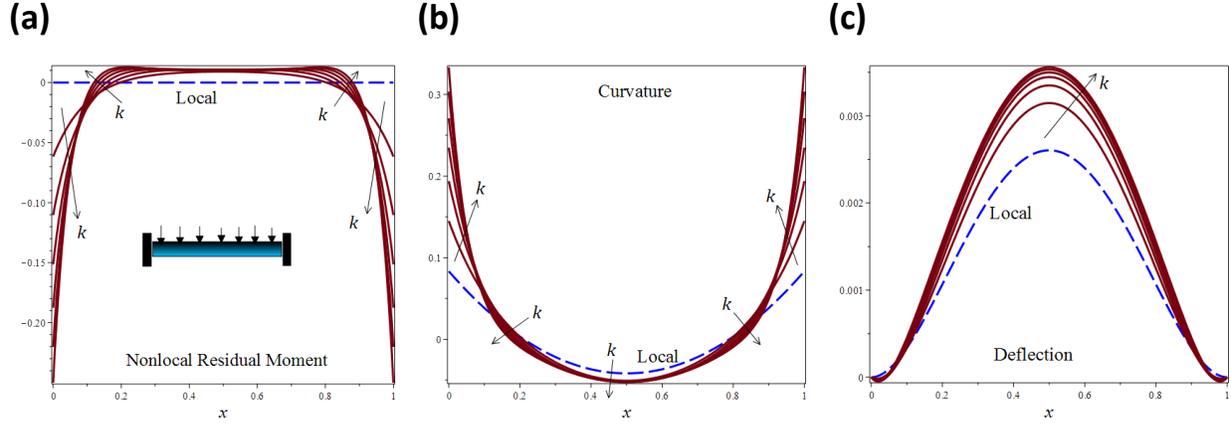

**Figure 4: Clamped-clamped beam under a uniform load ($q = 1$):** (a) The nonlocal residual moment ($\mathcal{M}(x)$), (b) the curvature ($\frac{d^2 w(x)}{dx^2} = \frac{M(x)}{(\lambda_0 + 2\mu_0)\frac{bh^3}{12}}$), and (c) the deflection ($w(x)$) at each iteration ($k$). The broken curves represent the local solutions. $\xi_\lambda = 1/3$, $\xi_\mu = 2/3$, $\ell_\lambda = \ell_\mu = 0.1$, $L = 1$.

The nonlocal residual moment ($\mathcal{M}(x)$) is zero in the first iteration to give the local solution. With the iterations, the nonlocal moment grows and converges to its complete value. This nonlocal residual moment adapt the boundary value problem to give updated solutions of the different fields of the beam. For the first time, nonlocal residual fields can be obtained by the proposed iterative nonlocal residual elasticity.

The obtained results in Figures 2-4 are consistent. All fields represented in these figures for the different beam configurations reflect the inherited softness of the nonlocal elasticity. The nonlocal curvatures and deflections are obtained larger than those of the local elasticity.

In addition, the obtained solutions fulfill the requirement of the global balance. The proposed iterative approach is operated based on fulfilling the equilibrium requirements globally in each iteration. This can be demonstrated by depicting the total nonlocal bending $M_{total}(x)$ (equation (4)), shear force $dM_{total}(x)/dx$, and loading $d^2 M_{total}(x)/dx^2$ fields in Figure 5 for the different beam configurations. For a cantilever beam with a length $L$ and under a point load of $F$, the global elastostatic equilibrium requires the total moment $M_{total}$ is zero at the free-end and increases linearly to $FL$ at the fixed end, and a constant shear force distribution over the beam. It is clear that the depicted results in Figure 5(a) perfectly match these requirements. In addition to the cantilever beam, the equilibrium requirements of the other beams are effectively fulfilled, as shown in Figures 5(b) and 5(c). Furthermore, the external loading distribution is obtained when differentiating the total moment twice ($d^2 M_{total}(x)/dx^2$), as shown in Figure 5. These observations demonstrate the effectiveness of the proposed iterative nonlocal residual elasticity.

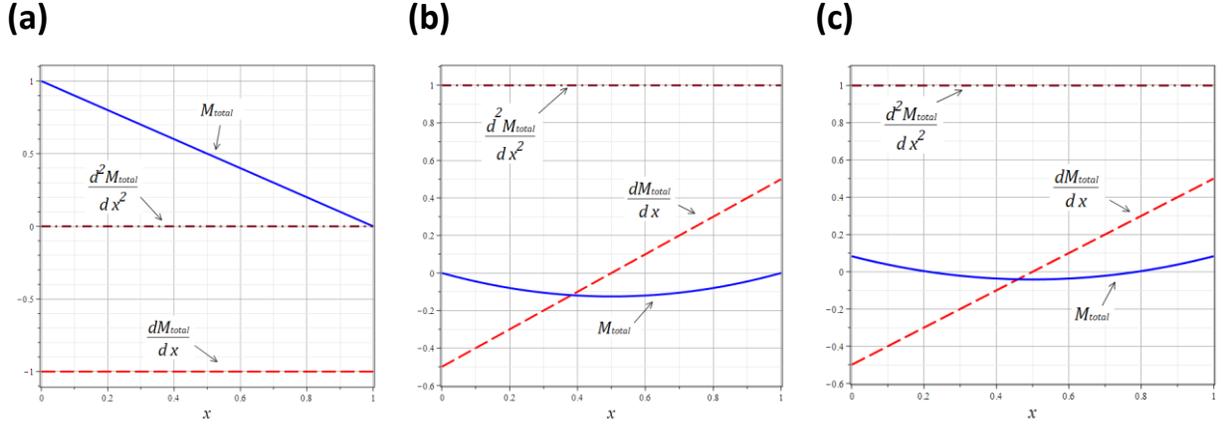

**Figure 5: Fulfillment of the global balance requirements:** The total nonlocal bending $M_{total}(x)$ (equation (4)), the shear force $dM_{total}(x)/dx$, and the loading $d^2M_{total}(x)/dx^2$ fields of (a) cantilever under a unit point load, (b) simply supported beam under a unit-uniform load, and (c) clamped-clamped beam under a unit-uniform load. ($L = 1$).

## Conclusions

In this study, a new nonlocal elasticity, which completely forms the nonlocal residual fields, was developed. This formulation was motivated by the existing complications of finding solutions of the conventional model of the nonlocal elasticity early proposed by Eringen. The new nonlocal elasticity was given the name "iterative nonlocal residual elasticity". This new nonlocal elasticity is centered upon expressing the dynamic equilibrium requirements based on a nonlocal residual stress field. The nonlocal residual stress was formulated to model the material stiffness due to non-neighbor interactions. Solutions of the boundary value problems formed based on the new nonlocal elasticity are determined using an iterative procedure. When the iterative procedure is implemented, the nonlocal boundary value problem is converted into a local boundary value problem of an elastic domain exposed to a residual field. This residual field is a nonlocal-type field, which grows with the iterations and is used to correct the local field problem for the nonlocal fields of the elastic domain. This iterative procedure is an integrated part of the new nonlocal elasticity.

The iterative nonlocal residual elasticity was applied to Euler-Bernoulli beams to verify the convergence to the targeted nonlocal solution and the fulfillment of the global balance requirements. It was shown that the iterative nonlocal residual elasticity decomposes the total nonlocal stress field into a local-type stress field and a nonlocal residual stress field. Thus, for the first time, information about nonlocal residual stress fields can be obtained using the proposed nonlocal elasticity. Findings presented in this study came to demonstrate that the complications of Eringen's nonlocal elasticity are effectively resolved when using the iterative nonlocal residual elasticity.

## Compliance with ethical standards

**Conflict of interest** The author declares that he has no conflicts of interest.